\documentclass[pre]{revtex4}

\usepackage[dvipdfm]{graphicx}

\usepackage{amsmath}
\usepackage{color}

\newcommand{\KILL}[1]{}

\usepackage{ulem}

\begin{document}

\title{Synchronization of uncoupled oscillators by common gamma
  impulses: from phase locking to noise-induced synchronization}

\date{\today}
 
\author{Shigefumi Hata$^{1}$, Takeaki Shimokawa$^{2}$, Kensuke
  Arai$^{3}$, and Hiroya Nakao$^{1,4}$}

\affiliation{
  $^{1}$Department of Physics, Kyoto University, Kyoto 606-8502,Japan\\
  $^{2}$ATR Neural Information Analysis Laboratories, 2-2-2 Hikaridai, Seika-cho, Soraku-gun, Kyoto 619-0288, Japan\\
  $^{3}$RIKEN Brain Science Institute, Wako, Saitama 351-0198, Japan\\
  $^{4}$JST, CREST, Kyoto 606-8502, Japan}

\begin{abstract}
  Nonlinear oscillators can mutually synchronize when they are driven
  by common external impulses.  Two important scenarios are (i)
  synchronization resulting from phase locking of each oscillator to
  regular periodic impulses and (ii) noise-induced synchronization
  caused by Poisson random impulses, but their difference has not been
  fully quantified.  Here we analyze a pair of uncoupled oscillators
  subject to common random impulses with gamma-distributed intervals,
  which can be smoothly interpolated between regular periodic and random Poisson
  impulses. Their dynamics are characterized by phase distributions, frequency detuning, Lyapunov exponents, and information-theoretic measures, which clearly reveal the differences between the two synchronization scenarios.
\end{abstract}


\maketitle


\section{Introduction}

Rhythmic elements are basic building blocks of Nature at human scales.
They play particularly important roles in the functioning of
biological systems, such as cardiac muscle cells, pacemaker neurons,
and animals or plants following circadian or circannual rhythms.
Recently, synchronization among non-interacting rhythmic elements
induced by common fluctuating external force has attracted much
attention in diverse fields.  It has been demonstrated for a wide
class of rhythmic elements ranging from lasers and electronic circuits
to biological systems~\cite{Uchida-Roy, Yoshida, Arai-Nakao,
  Nagai-Nakao, Mainen-Sejnowski, Binder-Powers, Galan,Danzl}.  For
example, reliable synchronous response of neurons due to common or
shared input signals has been actively discussed in
neuroscience~\cite{Ermentrout-Galan-Urban}.  In ecology, it is known
that, due to common climate fluctuations, populations of plants
exhibit large-scale synchronized flowering and production of seed
crops also fluctuate synchronously from year to year~\cite{Royama,
  Ranta, Koenig}, and such phenomena are generally termed the ``Moran
effect''~\cite{Moran}.

We often characterize a rhythmic element as a limit-cycle
oscillator. Populations of coupled limit-cycle oscillators show a
variety of interesting collective behavior, including mutual
synchronization, wave propagation, and chaos~\cite{Winfree, Kuramoto,
  Pikovsky, Glass}.  It is also well known that uncoupled limit-cycle
oscillators can mutually synchronize when they are driven by common
impulses~\cite{Winfree, Pikovsky, Glass, Yamanobe, Nakao-Arai,
  Arai-Nakao}.  The simplest classical situation is, of course,
synchronization due to phase locking of an oscillator to
common periodic impulses~\cite{Glass}.  Another interesting situation
is noise-induced synchronization as mentioned above, caused
e.g. by common random Poisson impulses~\cite{Pikovsky, Yamanobe,
  Nakao-Arai, Arai-Nakao}.  The oscillators can also synchronize when
the driving impulses have intermediate statistics between the periodic
and the Poisson impulses, as shown by Yamanobe {\it et
  al.}~\cite{Yamanobe} for a model of pacemaker neurons driven by
gamma-distributed impulses.

This prompts the question, what is the difference between the
synchronization due to phase locking and noise-induced
synchronization?  Though drive-response configuration of the impulse
source and the oscillators is the same for both cases, it is expected
that there should be some difference in their synchronization dynamics,
reflecting different characteristics of the driving impulses.
In this paper, we address this issue by considering uncoupled limit-cycle oscillators driven by gamma impulses, which can be smoothly interpolated
between regular periodic and random Poisson impulses.  We examine the
transition from phase locking to noise-induced synchronization
as the statistics of the gamma impulses is varied, and quantify their
difference using phase distributions, Lyapunov exponents, and
information-theoretic measures.

Effect of common gamma impulses on limit-cycle oscillators was previously treated in the paper by
Yamanobe {\it et al.}~\cite{Yamanobe}, but its main focus was not on
the transition between the two types of synchronization but rather on
physiologically realistic characterization of pacemaker neurons.  We
here conduct a systematic, quantitative analysis of the transition
between the two different synchronization dynamics.
We will reveal that the two types of synchronization dynamics are clearly different in many aspects, e.g. their stability, fluctuations, and statistical dependence on the driving impulses.

This paper is organized as follows.  In Sec.~II, we introduce a model
of uncoupled oscillators subject to common gamma impulses and
demonstrate its synchronization dynamics for different types of
driving impulses.  In Sec.~III, we analyze phase distributions of the
oscillators using Frobenius-Perron-type equations.  In Sec.~IV, we focus on frequency detuning of the oscillators due to impulsive driving.
In Sec.~V, we examine Lyapunov exponents and their fluctuations in the synchronized
states.  In Sec.~VI, we characterize mutual relationships among the
impulse source and the driven oscillators by utilizing
information-theoretic measures.
Section~VII summarizes the results.


\section{Limit-cycle oscillators driven by gamma impulses}

In this section, we introduce phase oscillators driven by common gamma
impulses and demonstrate their synchronization dynamics.

\subsection{Random phase map}

Based on the standard phase reduction theory of limit
cycles~\cite{Winfree, Kuramoto, Pikovsky, Nakao-Arai, Arai-Nakao}, we
can describe a limit-cycle oscillator using only its phase variable,
$\theta \in [0,1)$, defined along its stable limit cycle provided that
the intervals between the impulses are sufficiently longer than the
amplitude relaxation time of the oscillator state to the limit-cycle
orbit.  The key quantity for this description is the phase response
curve (PRC) $G(\theta)$ of the oscillator~\cite{Winfree}, which
encapsulates its dynamical properties.  $G(\theta)$ is a periodic
function on $[0,1)$ that gives asymptotic response of the oscillator
phase to an impulsive perturbation given at phase $\theta$.  It has
been measured experimentally, e.g., for repetitively firing
neurons~\cite{Tateno-Robinson, Galan-Ermentrout-Urban}.  When the
limit-cycle oscillator is kicked by an impulse at phase $\theta$, the
phase is mapped to a new value $F(\theta) = \theta + G(\theta)$.  This
function $F(\theta)$ is sometimes called the ``phase transition
curve'' in the literature.

The dynamics of the oscillator between two successive impulses
consists of a jump caused by the first impulse and subsequent free
rotation until the arrival of the second impulse~\cite{Nakao-Arai,
  Arai-Nakao, Arai-Nakao2}.  The oscillator is also subject to small
temporal fluctuations of various origins.  We assume that each oscillator
receives common driving impulses at times $\{ t_1, t_2, \cdots, t_n,
\cdots \}$, and denote the phase of the oscillator just before the
$n$th impulse by $\theta_n \in [0,1)$.  The effect of the $n$th
impulse is to map the oscillator phase $\theta_n$ to
$F(\theta_n)$.  To incorporate the effect of small fluctuations, we also
apply a weak additive independent noise $\eta_{n}$ before the mapping.  The phase
$\phi_{n}$ just after the $n$th impulse is thus given by $\phi_{n} =
F(\theta_{n} + \eta_{n})$.  After receiving the $n$th impulse, the
oscillator rotates with a constant frequency until $t_{n+1}$ at which
the $(n+1)$th impulse arrives.  Therefore, the phase $\theta_{n+1}$
just before the $(n+1)$th impulse is given by
\begin{equation}
  \theta_{n+1} = F(\theta_{n}+\eta_{n}) + \omega \tau_{n}
  \;
  \textrm{ mod }1,
  \label{eqRPM}
\end{equation}
where $\tau_{n} = t_{n+1} - t_{n}$ is the inter-impulse interval,
$\omega$ is the frequency of the oscillator, and we have taken modulo
$1$ of the phase to confine it in $[0,1)$.  Since $\tau_{n}$ and
$\eta_{n}$ are random variables, this equation gives a random map,
which we will call a ``random phase map'' hereafter. When
we consider multiple oscillators, $\tau_n$ is common to all
oscillators, whereas $\eta_n$ is different from oscillator to
oscillator.
Note that $\eta_{n}$ represents small temporal fluctuations of the dynamics of oscillators, but not static heterogeneities in their natural frequencies.  Mean frequency of the driven oscillator exhibits qualitatively different dependence on the driving impulse between phase locking and noise-induced synchronization, as we show in Sec. IV.

As the simplest and typical example, we assume that the PRC is given by a sinusoidal function $G(\theta) = c \sin (2\pi \theta)$ in the following numerical analysis, so that the phase map is given by
\begin{equation}
	F(\theta) = \theta + c \sin (2\pi \theta).
	\label{PM}
\end{equation}
The parameter $c$ controls the magnitude of the nonlinearity, which may also
be regarded as the intensity of the impulse.  Such a PRC with both
positive and negative lobes are called Type-II~\cite{Hansel,
  Abouzeid-Ermentrout, Marella-Ermentrout, Danzl}, which generally
arises near the Hopf bifurcation of limit-cycle
oscillators~\cite{Hansel, Abouzeid-Ermentrout, Marella-Ermentrout,
  Izhikevich, Arai-Nakao, Brown}.  Another typical example is the
Type-I PRC with a single positive lobe, e.g. $G(\theta) = c \{ 1 -
\cos( 2 \pi \theta ) \}$, which generally arises near the SNIC
(saddle-node on invariant circle) bifurcation of
oscillators~\cite{Ermentrout-typeI, Brown, Marella-Ermentrout, Danzl}.
However, difference between these two PRCs can roughly be eliminated by simply shifting the frequency $\omega$ and the origin of the phase $\theta$ in the present setup, and qualitatively similar results are expected for both PRCs.  Actually, the two types of PRCs yielded almost the same results in our numerical analysis. 
Thus, the type-II PRC, Eq.~(\ref{PM}), gives us general insights and we only present the results for this case hereafter.


\begin{figure}[tbhp]
  \begin{center}
   \includegraphics[width=0.75\hsize]{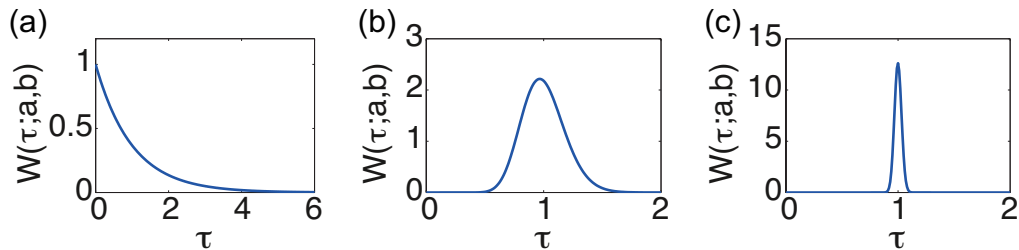}
   \caption{(Color online) The gamma distribution $W(\tau;a,b)$ for
     (a) $a=1, b=1$, (b) $a=30, b=1/30$, and (c) $a=1000,
     b=1/1000$. The mean value is fixed at $\langle \tau \rangle = ab
     = 1$.}
    \label{fig.1}
  \end{center}
\end{figure}

\subsection{Gamma impulses}

We consider driving impulses that have ``intermediate'' statistics
between regular periodic impulses and random Poisson impulses. As
such impulses, we adopt the gamma impulses~\cite{Yamanobe}, whose
inter-impulse interval $\tau$ obeys the gamma
distribution~\cite{Yamanobe, Shimokawa},
\begin{equation}
  W(\tau ; a, b)=\tau^{a-1} \frac{e^{-\frac{\tau}{b}}}{\Gamma(a)b^a},\label{eq:gamma}
\end{equation}
where $a$ and $b$ are real positive parameters.  The mean interval is
given by $\langle \tau \rangle= \int_{0}^{\infty} W(\tau ; a,b) \tau
d\tau = ab$.  The gamma distribution has the following properties.
Firstly, when $a=1$, $W(\tau; a,b)$ gives an exponential distribution,
\begin{equation}
  W(\tau ; a=1, b) = \frac{e^{-\frac{\tau}{b}}}{b},
\end{equation}
which means that the impulses obey the standard Poisson random
process.  Secondly, by taking the limit $a \rightarrow \infty$ and $b
\to 0$ with $\langle \tau \rangle = ab$ fixed, the gamma distribution
approaches the delta function,
\begin{equation}
  W(\tau ; a, b) \overset{a\rightarrow \infty, b \to 0}{\longrightarrow}
  \delta(\tau-\langle \tau \rangle).
\end{equation}
In this limit, the inter-impulse interval is always $\langle \tau
\rangle$, so that the impulses become periodic.

Thus, the parameter $a$, which we call a shape parameter, determines
the property of the gamma impulses as shown in Fig.~\ref{fig.1}.  By
varying the value of $a$ between $1$ and $\infty$ with $\langle \tau
\rangle = ab$ fixed, the gamma impulses can exhibit intermediate
properties between random Poisson and regular periodic impulses while
keeping the same mean inter-impulse interval.  We examine uncoupled
oscillators driven by gamma impulses and see how synchronization
dynamics of the oscillators depends on the shape parameter $a$ in the
following.

The gamma distribution has several nice mathematical properties and naturally arises under a few simple assumptions when the mean and the irregularity of impulse sequences, which respectively correspond to the mean interval $\langle \tau \rangle$ and the shape parameter $a$~\cite{Shimokawa}, are given. We thus use the gamma impulses in the present paper, though alternative approaches for varying the regularity of the driving signal should also be possible, e.g., by using chaotic dynamical systems~\cite{Santos}.

\subsection{Synchronization by common driving impulses}

We here demonstrate synchronization of uncoupled oscillators induced
by common impulses for the periodic, Poisson, and intermediate cases
by direct numerical simulations.  We fix the magnitude of nonlinearity  (or the intensity of impulses) of Eq.~(\ref{PM}) at $c=0.1$ throughout our numerical simulations, which is sufficiently small and therefore the map is not chaotic even for strictly periodic impulses.  This is because we are focusing on the synchronization dynamics of limit cycles (note that the sinusoidal Type II phase map with periodic impulses is nothing but the well-known circle map~\cite{Pikovsky}).  Strong Poisson impulses may also yield positive Lyapunov exponents and lead to desynchronization of oscillators as shown in Ref.~\cite{Arai-Nakao}, but we do not consider such a situation in the present paper.

\begin{figure}[tbhp]
  \begin{center}
    \includegraphics[width=0.3\hsize]{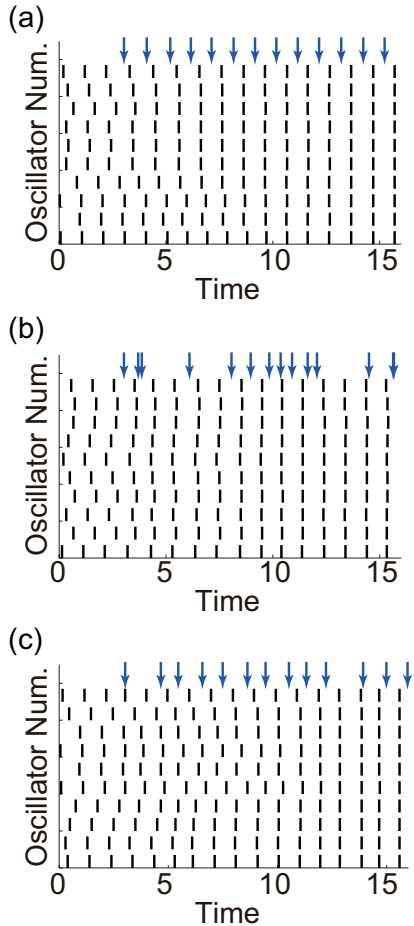}
    \caption{(Color online) Synchronization of uncoupled oscillators
      induced by common impulses.  Rasters show zero-crossing events
      of individual oscillators.  Each arrow shows the time when an
      impulse arrives.  (a) Phase locking ($a=1000$). (b)
      Noise-induced synchronization ($a=1$). (c) Synchronization
      induced by gamma impulses ($a=30$).}
    \label{fig.2}
  \end{center}
\end{figure}

Figure \ref{fig.2} shows typical synchronization dynamics of uncoupled
oscillators induced by three types of common impulses, where
zero-crossing points of $10$ oscillators are shown in raster plots.
The mean interval of the impulses is set at $\langle \tau \rangle = ab
= 1$.  The period of the oscillator is also taken as $T = 1/\omega =
1$ and is thus equals to the mean interval.

(i) Phase locking to periodic impulses [Fig.~\ref{fig.2}.(a)].  If the
impulses are periodic and their intervals are nearly equal to (or more
generally rational multiples of) the intrinsic rotation period of the
oscillators, namely, if they are resonant, each oscillator becomes
phase locked to the impulses.  As a consequence, uncoupled oscillators
driven by common periodic impulses synchronize with each other.

(ii) Noise-induced synchronization by Poisson impulses
[Fig.~\ref{fig.2}.(b)].  The oscillators also synchronize when they
are driven by common Poisson impulses of appropriate intensity.  It
has been theoretically and experimentally shown that uncoupled
limit-cycle oscillators subject to weak common Poisson impulses
generally synchronize with each other, irrespective of their
details~\cite{Pikovsky, Nakao-Arai, Arai-Nakao}.

(iii) Synchronization induced by gamma impulses
[Fig.~\ref{fig.2}.(c)].  Uncoupled oscillators subject to the gamma
impulses with intermediate statistics between the periodic and Poisson
impulses can also synchronize mutually~\cite{Yamanobe}.

Thus, the uncoupled oscillators synchronized with each other by the effect of the common impulses for all values of the shape parameter.  It is however not easy to catch the quantitative differences in their synchronization dynamics just from these figures.  In the following sections, we will characterize the differences for varying types of common impulses using phase distributions, frequency detuning, Lyapunov exponents, and information-theoretic measures.


\section{Phase distributions}

In this section, we introduce Frobenius-Perron-type
equations~\cite{Lasota-Mackey, Doi, Nakao-Arai} that describe
evolution of phase distributions.  Using them, we examine how the
relations among the oscillator phases and the inter-impulse intervals
depend on the shape parameter, which reveal differences between phase locking and noise-induced synchronization.

\subsection{Single-oscillator Frobenius-Perron equation}

Let us consider a single-oscillator probability density function (PDF)
$P_{n}(\theta)$ of the phase $\theta_{n}$ at time step $n$, just
before the $n$th impulse arrives. To obtain a Frobenius-Perron
equation describing the dynamics of $P_n(\theta)$, it is convenient to
consider a PDF $P_n(\bar{\theta})$ of the ``unwrapped'' phase
$\bar{\theta}_n$ defined in $(-\infty, \infty)$, which obeys the
following random phase map without the modulo~$1$,
\begin{equation}
  \bar{\theta}_{n+1} = F(\theta_{n}+\eta_{n}) + \omega \tau_{n}.
  \label{eq:RPM2}
\end{equation}
The PDF $P_{n+1}(\bar{\theta})$ of $\bar{\theta}_{n+1}$ is given by
the following Frobenius-Perron equation~\cite{Lasota-Mackey,Ott}:
\begin{align}
  P_{n+1}( \bar{\theta}) = \int_0^1 d\theta \int_0^{\infty} d\tau \int
  _{-\infty}^{\infty} d\eta P_{n}(\theta) W(\tau ) R(\eta) \delta
  \left( \bar{\theta} - F( \theta + \eta ) - \omega \tau
  \right), 
  \label{eq:03}
\end{align}
where $W(\tau)$ is the PDF of the inter-impulse intervals, namely, the
gamma distribution given in Eq.~(\ref{eq:gamma}), and
\begin{align}
  R(\eta) = \frac{1}{\sqrt{2 \pi \sigma^2}} \exp\left( -
    \frac{\eta^2}{2\sigma^2} \right)
\end{align}
is the probability density function of the independent additive noise,
which we assume to be zero-mean Gaussian of variance $\sigma^2$.  The
probability density function of the wrapped phase, $\theta =
\bar{\theta} \mbox{ mod } 1 \in [0,1)$, can be calculated
as~\cite{Ermentrout-Saunders}
\begin{equation}
  P_{n}(\theta) = \sum_{k=-\infty}^{\infty} P_{n}(\bar{\theta} + k ).
\end{equation}
Thus, from Eq.~(\ref{eq:03}), the Frobenius-Perron equation describing
$P_n(\theta)$ is obtained as
\begin{align}
  P_{n+1}(\theta) = \sum_{k=-\infty}^{\infty} \int_0^1 d\theta
  \int_0^{\infty} d\tau \int_{-\infty}^{\infty} d\eta P_{n}(\theta)
  W(\tau) R(\eta) \delta\left (\theta - F(\theta + \eta) - \omega \tau
    - k \right ),
  \label{eq:FP1}
\end{align}
which describes the single-oscillator PDF $P_{n}(\theta)$ of the phase
$\theta_{n}$ just before the $n$th impulse.

Stationary solutions of Eq.~(\ref{eq:FP1}) gives the PDF of the
oscillator phase driven by gamma impulses and additive noise in
statistically steady states.  In Ref.~\cite{Nakao-Arai}, we have
treated a similar Frobenius-Perron equation perturbatively to obtain
the stationary phase PDF of the oscillator under weak Poisson impulses
and calculated the Lyapunov exponents.  In Ref.~\cite{Doi}, Doi {\it
  et al.} analyzed the spectral properties of a similar Frobenius-Perron equation
(driven by periodic impulses) to characterize noisy phase locking of
a limit-cycle oscillator.

\subsection{Two-oscillator Frobenius-Perron equation}

To investigate phase relationships between a pair of oscillators
(denoted here as $A$ and $B$) subject to common impulses, we introduce
a joint PDF $P_n(\theta^{A}, \theta^{B})$ of their phases
$\theta^{A}_n$ and $\theta^{B}_n$ just before the $n$th impulse. The dynamics of the pair is given by
\begin{align}
  \theta^{A}_{n+1} &= F(\theta^{A}_{n}+\eta^{A}_{n}) + \omega \tau_{n}
  \;
  \textrm{ mod }1, \cr
  \theta^{B}_{n+1} &= F(\theta^{B}_{n}+\eta^{B}_{n}) + \omega \tau_{n}
  \;
  \textrm{ mod }1,
\end{align}
where $\eta_n^A$ and $\eta_n^B$ are independent additive noise terms
($\tau_n$ is common to both oscillators).  Similarly to the
single-oscillator case, we can derive a Frobenius-Perron equation describing the two-oscillator phase PDF $P_n(\theta^{A}, \theta^{B})$
as
\begin{align}
  P_{n+1}(\theta^{A}, \theta^{B}) = \sum_{k_A=-\infty}^{\infty}
  \sum_{k_B=-\infty}^{\infty} \int_0^1 d\theta^{A'} \int_0^1
  d\theta^{B'} \int_0^{\infty} d\tau \int_{-\infty}^{\infty}
  d\eta^{A'} \int_{-\infty}^{\infty} d\eta^{B'} P_{n}(\theta^{A'},
  \theta^{B'}) W(\tau) \times\cr R(\eta^{A'}) R(\eta^{B'}) \delta\left
    ( \theta^{A} - F(\theta^{A'} + \eta^{A'}) - \omega \tau -
    k_A\right ) \delta\left ( \theta^{B} - F(\theta^{B'} + \eta^{B'})
    - \omega \tau - k_B\right ).
  \label{eq:FP2}
\end{align}
The stationary solution of Eq.~(\ref{eq:FP2}) gives a PDF of the pair
of oscillator phases driven by common impulses and independent
additive noise in the statistically steady state.

In Ref.~\cite{Arai-Nakao2}, we have analyzed a similar two-oscillator
Frobenius-Perron equation by averaging out the fast dynamics of the
mean phase to obtain an approximate Frobenius-Perron equation
describing only the phase difference of two oscillators driven by
common Poisson impulses.  Since we are interested in pair phase
distributions, not only in the distribution of phase differences, we
directly solve Eq.~(\ref{eq:FP2}) numerically for gamma impulses in the following.


\begin{figure}[tbhp]
  \begin{center}
   \includegraphics[width=0.75\hsize]{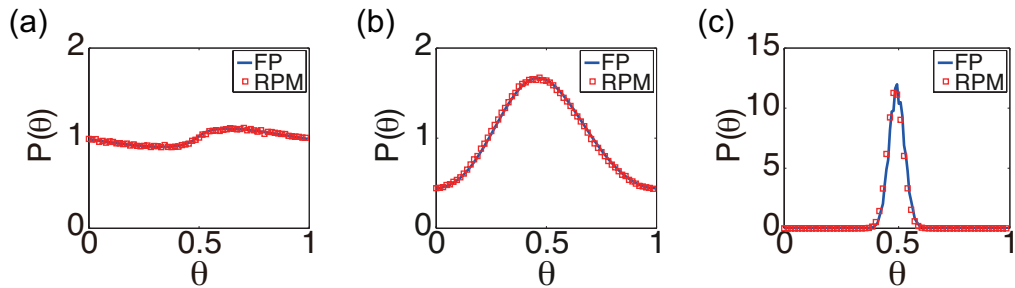}
   \caption{(Color online) Stationary single-oscillator probability
     density function $P(\theta)$ of the phase $\theta$ just before
     the arrival of each impulse.  Blue curves are obtained by numerically solving the
     Frobenius-Perron equation and red squares are the results of
     direct numerical simulations of the random phase map.  (a)
     Noise-induced synchronization ($a=1$). (b) Synchronization
     induced by gamma impulses ($a=30$). (c) Phase locking
     ($a=1000$).}
    \label{fig.3}
  \end{center}
\end{figure}


\begin{figure}[tbhp]
  \begin{center}
    \includegraphics[width=0.75\hsize]{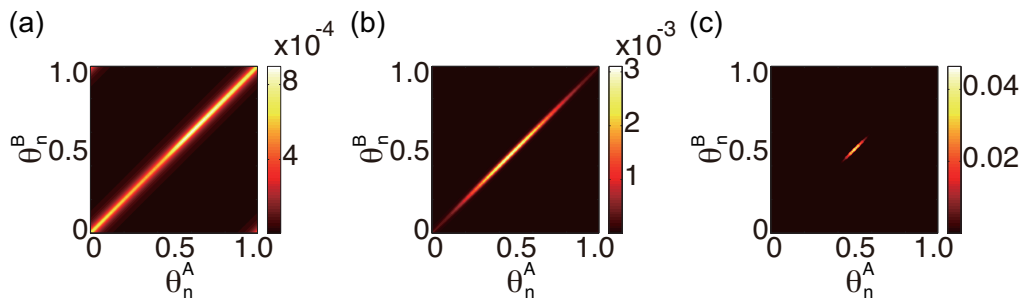}
    \caption{(Color online) Stationary two-oscillator joint
      probability density function $P(\theta^A, \theta^B)$ of the
      phase pair $\theta^A$ and $\theta^B$ just before the impulses.
      (a) Noise-induces synchronization ($a=1$). (b) Synchronization
      induced by intermediate impulses ($a=30$). (c) Phase locking
      ($a=1000$).}
    \label{fig.4}
  \end{center}
\end{figure}


\begin{figure}[tbhp]
  \begin{center}
    \includegraphics[width=0.75\hsize]{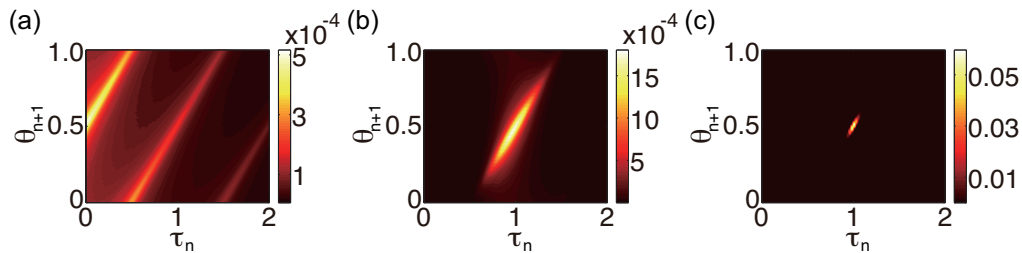}
    \caption{(Color online) Stationary joint probability distribution
      $P(\tau, \theta)$ of the inter-impulse interval $\tau$ and the
      phase $\theta$ immediately after this interval. (a)
      Noise-induced synchronization ($a=1$). (b) Synchronization
      induced by intermediate impulses ($a=30$). (c) Phase locking
      ($a=1000$).  }
    \label{fig.5} 
  \end{center}
\end{figure}

\subsection{Numerical results}

We here present stationary phase PDFs obtained by numerically solving
the Frobenius-Perron equations.  We set the frequency of each
oscillator at $\omega = 1$.  The period of the oscillator is $T =
1/\omega = 1$ and is equal to the mean period of impulses $\langle
\tau \rangle = 1$.  Thus, when the shape parameter $a$ is sufficiently
large and the impulses are nearly periodic, phase locking should
occur.  In the numerical analysis, each variable is discretized using
$100$-$200$ grid points.  The calculation cost can be drastically
reduced by devising numerical algorithms that use Fourier
representation in calculating convolutions of the Frobenius-Perron
equations (see Appendix for details).

\subsubsection{Single-oscillator phase distributions}

Figures \ref{fig.3}(a)-(c) show the stationary PDF $P(\theta)$ of a
single oscillator phase $\theta$ just before each impulse for nearly
periodic ($a=1000$), intermediate ($a=30$), and Poisson ($a=1$) cases.
Numerical solutions of the Frobenius-Perron equation~(\ref{eq:FP1}) and the results of direct numerical simulations of the random phase map~(\ref{eqRPM}) agree nicely.
The intensity of the weak independent Gaussian noise is $\sigma =
0.01$.  In all cases, the oscillators synchronize with each other (to
the independent noise level).  However, depending on the shape
parameter $a$, the phase PDF $P(\theta)$ significantly differs.  For
nearly periodic impulses, $P(\theta)$ has a sharp peak, indicating
that the oscillator phase is actually entrained to the driving impulse
with some fixed phase shift.  In contrast, $P(\theta)$ is roughly
uniformly distributed for Poisson impulses, which means that the
relationship between the oscillator phase and the impulse timing is not fixed and hence not entrained.
In the intermediate case, $P(\theta)$ has a broad but still clear one-humped shape, implying that the phase is not completely entrained to the impulses but still possesses a certain level of coherence with respect to the driving impulses.

\subsubsection{Two-oscillator phase distributions}

Figures \ref{fig.4}(a)-(c) show the stationary joint phase PDF
$P(\theta^{A},\theta^{B})$ of a pair of oscillators for nearly
periodic ($a=1000$), intermediate ($a=30$), and Poisson ($a=1$)
impulses obtained by numerically solving the Frobenius-Perron equation~(\ref{eq:FP2}).
In all cases, the oscillators synchronize with each other,
so that their phases are distributed along the diagonal line,
$\theta^{A} = \theta^{B}$.  However, their distribution strongly
depends on the shape parameter $a$.  For nearly periodic impulses
($a=1000$), the oscillator is almost phase-locked to the impulses and
thus the PDF has a sharp peak near the center, indicating that the
oscillators are not only mutually synchronized but both of them are
entrained to the impulses.  As the parameter $a$ decreases, the
distribution becomes broader, and for Poisson impulses ($a=1$), the
phases $\theta^{A}$ and $\theta^{B}$ are broadly distributed along the
diagonal line, indicating that they are synchronized but not entrained
by the impulses anymore.

\subsubsection{Joint distributions of the inter-impulse intervals and the oscillator phases}

To analyze how the driving impulses affect the oscillator phase, we
also calculate the joint PDF of the inter-impulse interval and the
oscillator phase just after this interval in the statistically steady
states.
The stationary joint PDFs $P(\tau, \theta) = P(\theta | \tau) W(\tau)$ of the impulse interval $\tau$ and the phase
$\theta$ observed immediately after this interval are calculated from the Frobenius-Perron equation.
Figures \ref{fig.5}(a)-(c) show stationary joint PDFs
$P(\tau, \theta)$ for nearly periodic
($a=1000$), intermediate ($a=30$), and Poisson ($a=1$) cases. The distribution has a sharp peak in the nearly phase-locked
case ($a=1000$), indicating that the inter-impulse interval and the
oscillator phase just after this interval have almost a one-to-one
correspondence, namely, the oscillator phase is almost entrained by
the impulses.  As we decrease the shape parameter $a$, the
distribution gradually elongates, and in the Poisson limit ($a=1$),
such clear correspondence is lost (but they still retain a certain
degree of correlation).

\vspace{12pt}
Thus, all phase distributions clearly capture the essential difference between the phase locking and the noise-induced synchronization.  The relation among the oscillator phases and the inter-impulse intervals significantly differs depending on the shape parameter $a$, even though the oscillators themselves are always mutually synchronized.  As $a$ is increased, the oscillators tend to be more strictly phase locked to the driving impulses, whereas for smaller $a$, their dependence becomes weaker.  This observation will be quantified by information-theoretic measures in Sec. VI.


\section{Frequency detuning}

In this section, we consider how the dynamics of the oscillator is modulated by the driving impulses.  Specifically, we examine the frequency detuning~\cite{Pikovsky}, i.e., the difference between the mean frequency of the driven oscillator and that of the driving impulses, for varying values of the shape parameter.
This analysis will reveal another clear difference between the two types of synchronization scenarios.

\subsection{Mean frequency of the driven oscillator}
 
From the phase PDF $P(\theta)$ and the PRC $G(\theta)$, we can estimate the mean frequency of the impulse-driven oscillator as follows.
Using the random phase map for unwrapped phase $\bar{\theta}_n$, Eq.~(\ref{eq:RPM2}), the oscillator phase just before the $N$th impulse is given by
\begin{align}
	\bar{\theta}_N
	&= \bar{\theta_1} + \sum_{i=1}^{N-1}\left ( G(\bar{\theta_i} + \eta_{i})+\omega\tau_i \right ).
\end{align}
Therefore, long-time mean frequency $\omega'$ of the driven oscillator can be calculated from this equation as
\begin{align}
	\omega' 
	&= \lim_{N \to \infty} \frac{\bar{\theta}_N-\bar{\theta}_1}{N-1} \cr
	&= \lim_{N \to \infty} \left( \frac{1}{N-1} \sum_{i=1}^{N-1} G(\bar{\theta_{i}}  + \eta_{i}) + \frac{1}{N-1} \sum_{i=1}^{N-1} \omega \tau_{i} \right) \cr
	&= \int_0^1d\theta P(\theta)G(\theta) + \omega \int_0^\infty d\tau W(\tau) \cr
  & = \int_0^1d\theta P(\theta)G(\theta) + \omega,
\end{align}
where we have replaced the long-time average by a statistical average and used the periodicity of the PRC $G(\theta)$ and the normalization condition $\int_{0}^{\infty} d\tau W(\tau) = 1$.  Thus, the mean frequency detuning between the oscillator and the impulses is given by
\begin{equation}
	\omega ' -\Omega = \int_0^1d\theta P(\theta)G(\theta) + \omega - \Omega,
	\label{eq:FreqDet}
\end{equation}
where $\Omega = 1/\langle \tau \rangle$ is the mean frequency of the driving impulses.

\subsection{Numerical results}

\begin{figure}[tbhp]
  \begin{center}
    \includegraphics[width=0.5\hsize]{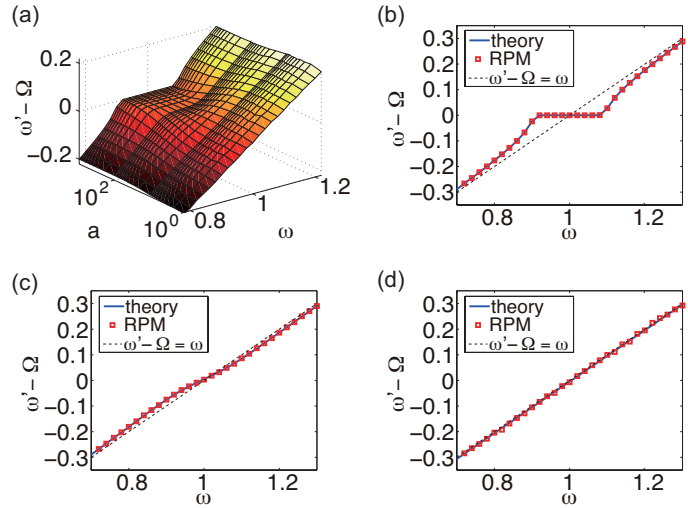}
    \caption{(Color online) Frequency detuning.
    (a) Dependence of the frequency detuning $\omega'-\Omega$ on the shape parameter $a$ and on the natural frequency $\omega$ of the driven oscillator, measured by direct numerical simulations of the random phase map.
    (b)-(d) Dependence of the frequency detuning on the natural frequency $\omega$ of oscillators for several values of the shape parameter $a$.  Blue solid lines are the results of the integral calculation, Eq.(\ref{eq:FreqDet}), and red squares are obtained by direct numerical simulations of the random phase map.
    (b) Phase locking ($a=1000$).
    (c) Synchronization induced by gamma impulses ($a=30$).
    (d) Noise-induced synchronization ($a=1$).}
    \label{fig.6}
  \end{center}
\end{figure}

We fix the mean frequency of the driving impulses at $\Omega = 1 / \langle \tau \rangle = 1$ and vary the shape parameter $a$ and the natural frequency $\omega$ of the oscillator in the numerical simulations.
Figure~\ref{fig.6}(a) plots the frequency detuning $\omega' - \Omega$ on the $(a, \omega)$ plane. 
For large $a$ and nearly resonant natural frequency, $\omega \simeq \Omega = 1$, the conditions for phase locking are satisfied and therefore the mean frequency $\omega'$ of the oscillator is locked to that of the driving impulses $\Omega$, yielding a plateau in the figure.
In contrast, for small $a$, the mean frequency $\omega'$ never locks to $\Omega$.

Figure~\ref{fig.6}(b) plots $\omega' - \Omega$ as a function of $\omega$ at $a=1000$, i.e., for approximately periodic impulses.  Results of direct numerical simulations agree well with the theoretical estimate, Eq.~(\ref{eq:FreqDet}), with the PDF $P(\theta)$ obtained numerically from the Frobenius-Perron equation.  A plateau satisfying $\omega' = \Omega$ can clearly be seen near $\omega = \Omega$.
In the intermediate case ($a=30$) shown in Fig.~\ref{fig.6}(c), clear phase-locking region no longer exist, however, the frequency of the oscillator is modulated by the driving impulses.
In the Poisson case ($a=1$), the mean frequency of the driven oscillator is not affected by the driving impulse as shown in Fig.~\ref{fig.6}(d). 

This observation provides us with further evidence on the difference between the two synchronization scenarios.  Actually, as pointed out by Yoshimura {\it et al.} in Ref.~\cite{Yoshimura-Arai-Davis}, absence of frequency locking between the oscillators is characteristic to common-noise-induced synchronization, in sharp contrast to ordinary phase locking due to mutual coupling (though they considered weak Gaussian noise instead of random impulses in their work).


\section{Stability of synchronized states}

In this section, to characterize the synchronization dynamics, we focus on Lyapunov exponents and their fluctuations in the synchronized states.  As we will see, differences between synchronization scenarios are well characterized by the fluctuations of the Lyapunov exponent.

\subsection{Lyapunov exponents and its variance}

To quantify statistical linear stability of the synchronized states,
we calculate the Lyapunov exponent and its variance.  Let us consider
a pair of oscillators and denote their phases at time $n$ as
$\theta_{n}$ and $\theta_{n}'$, respectively.  Linearized evolution of
a small phase difference $\Delta_{n} = \theta_{n}' - \theta_{n}$ is
given by
\begin{align}
  \Delta_{n+1}= & F'(\theta_n + \eta_n) \Delta_n,
\end{align}
where $F'(\theta) = d F(\theta) / d\theta$ is the instantaneous linear
growth rate of the phase difference.  The deviation $\Delta_{N}$ at
large time step $N$ is thus given by
\begin{align}
\left| \frac{\Delta_{N}}{\Delta_{0}} \right|
	= &
	\prod_{n=0}^{N-1} |F'(\theta_n + \eta_n)| 
	\simeq \exp[ N \langle \Lambda \rangle ],
\end{align}
where we have introduced the Lyapunov exponent $\langle \Lambda
\rangle$, defined by a long-time average of the linear growth rates as
\begin{align}
  \langle \Lambda \rangle =& \lim_{N \to \infty} \frac{1}{N} \ln\left|
    \frac{\Delta_{N}}{\Delta_{0}} \right| = \lim_{N \to \infty}
  \dfrac{1}{N}\sum_{n=0}^{N-1} \ln \left|
    F'\left(\theta_{n}+\eta_{n}\right) \right| \cr = & \int_0^1
  d\theta \int_{-\infty}^{\infty}d\eta P(\theta)R(\eta)\ln \left|
    F'\left(\theta+\eta\right) \right|.
\end{align}
In the last expression, we have replaced the long-time average by a
statistical average over the stationary phase PDF $P(\theta)$ and the
PDF $R(\eta)$ of the independent noise. The synchronized state is
stable if $\langle \Lambda \rangle$ is negative.

Similarly, variance of the linear growth rates, $\textrm{var}(\Lambda)
= \langle \Lambda^2 \rangle - \langle \Lambda \rangle^2$, can be
calculated as
\begin{align}
  \textrm{var}(\Lambda) =& \int_0^1 d\theta \int_{-\infty}^{\infty}
  d\eta P(\theta) R(\eta) \big\{ \ln \left| F'\left(\theta+\eta\right)
  \right| \big\}^2 \cr &- \left\{ \int_0^1 d\theta
    \int_{-\infty}^{\infty} d\eta P(\theta) R(\eta)\ln \left| F'\left(
        \theta + \eta \right) \right| \right\}^2.
\end{align}
As we will see, even if the Lyapunov exponent $\langle \Lambda
\rangle$ takes the same value, its variance $\textrm{var}(\Lambda)$
can significantly differ depending on the shape parameter $a$.


\begin{figure}[tbhp]
  \begin{center}
    \includegraphics[width=0.75\hsize]{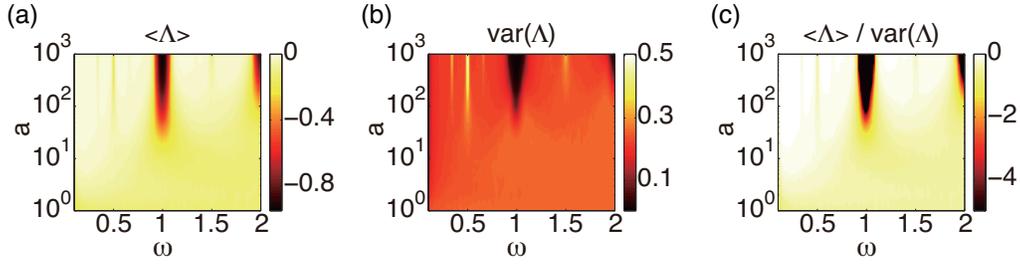}
    \caption{(Color online) Dependence of the Lyapunov exponent $\Lambda$ on the shape parameter $a$ and on the
      frequency $\omega$ of oscillators. (a) Mean Lyapunov exponent $\langle \Lambda \rangle$. (b) Variance of Lyapunov exponent $\textrm{var}(\Lambda)$.  (c) Ratio of the mean to the variance, $\langle \Lambda \rangle / \textrm{var}(\Lambda)$.}
    \label{fig.7}
  \end{center}
\end{figure}


\begin{figure}[tbhp]
  \begin{center}
    \includegraphics[width=0.75\hsize]{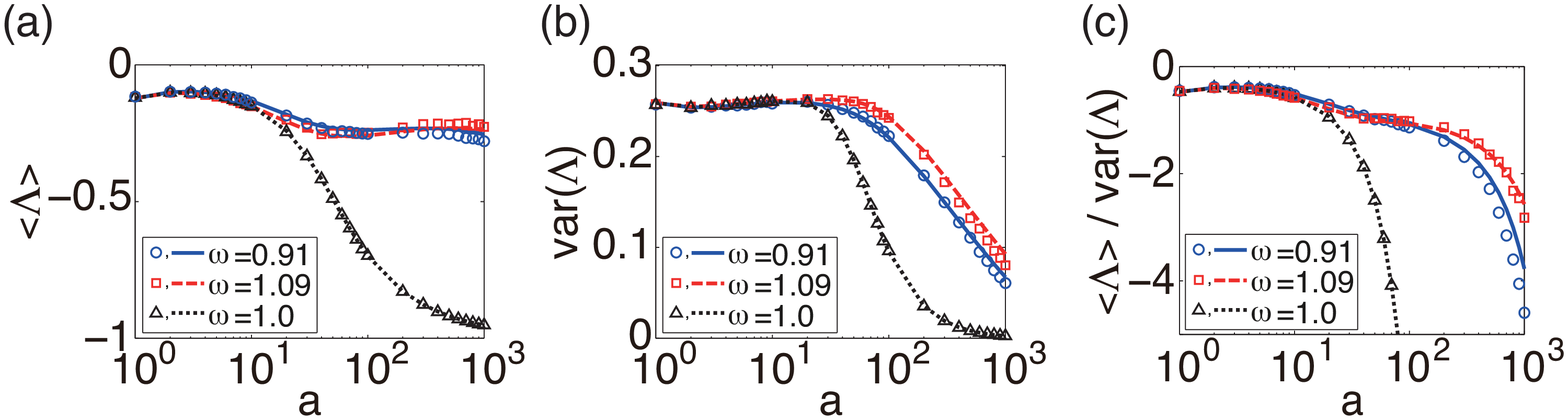}
    \caption{(Color online) Dependence of the mean and variance of
      Lyapunov exponent $\Lambda$ on the parameter $a$. The curves are calculated from the Frobenius-Perron equation and the symbols are the results of direct numerical simulations of the random phase map.
      (a) Mean $\langle \Lambda \rangle$. (b) Variance
      $\textrm{var}(\Lambda)$.}
    \label{fig.8}
  \end{center}
\end{figure}

\subsection{Numerical results}

Figure~\ref{fig.7}(a) plots the Lyapunov exponent $\langle \Lambda
\rangle$ on the $(a, \omega)$ plane, showing its dependence on the
shape parameter $a$ and on the natural frequency of the oscillator $\omega$.  The
mean period of the impulses is fixed at $\langle \tau \rangle = 1$.
For sufficiently large $a$, there exists bell-shaped regions near the resonant frequencies (i.e., $\omega = 1$ and $\omega = 2$), where the Lyapunov exponent takes large negative values.  These regions correspond to nearly phase-locked states and may be seen as a kind of Arnold tongues.  The Lyapunov exponent also takes small negative values outside this region, which corresponds to the noise-induced synchronized states.  Note that phase locking occurs only when the oscillator frequency is approximately resonant to the impulses, whereas the noise-induced synchronization occurs almost everywhere.

Dependence of the Lyapunov exponent $\langle \Lambda\rangle$ on the
shape parameter $a$ for several values of $\omega$ is plotted in
Fig.~\ref{fig.8}(a).  When the natural frequency is resonant to the mean inter-impulse interval ($\omega = \Omega = 1 / \langle \tau \rangle = 1$),
$\langle \Lambda \rangle$ smoothly decreases as we move from the
noise-induced synchronization ($a=1$) to the phase-locking ($a=1000$),
indicating that the phase locking is more stable than the
noise-induced synchronization.  However, for oscillator frequencies
near the edges of the phase-locking region ($\omega = 0.91$ and
$1.09$), $\langle \Lambda \rangle$ differs only little between $a=1$
and $a=1000$, so that the noise-induced synchronization can be
nearly stable as the phase locking.  Thus, though the Lyapunov
exponent $\langle \Lambda \rangle$ reflects the synchronization
dynamics, difference between the two types of synchronization cannot
be fully characterized just by looking at $\langle \Lambda \rangle$.

Figure~\ref{fig.7}(b) shows dependence of the variance of the Lyapunov exponent
exponent $\textrm{var}(\Lambda)$ on the shape parameter $a$ and on the oscillator frequency $\omega$,
and Figure~\ref{fig.8}(b) shows the dependence on the shape parameter $a$ for
$\omega = 0.91, 1$, and $1.09$.  In contrast to the Lyapunov exponent
itself, the variance $\textrm{var}(\Lambda)$ always decreases
considerably as $a$ is increased from the noise-induced
synchronization ($a=1$) to the phase-locking ($a=1000$), because the
phase PDF becomes much broader for smaller $a$ as we have seen in
Fig.~\ref{fig.3}.  Thus, the difference between the phase locking and
the noise-induced synchronization can be captured more clearly by the
fluctuations in the Lyapunov exponent, rather than by its long-time average.

The difference can also be clearly visualized by plotting the ratio of the Lyapunov exponent and its variance $\langle \Lambda \rangle / \textrm{var}(\Lambda)$ as in Fig.~\ref{fig.7}(c) and Fig.~\ref{fig.8}(c).  Actually, this ratio characterizes the intermittent dynamics of the phase differences between the oscillators as we explain below.

\subsection{Intermittent dynamics of phase differences}

\begin{figure}[tbhp]
  \begin{center}
    \includegraphics[width=0.3\hsize]{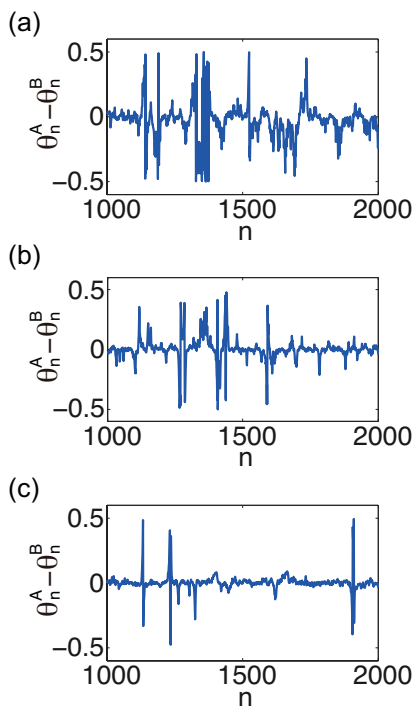}
    \caption{(Color online) Dynamics of the phase difference
      $\Delta_n$ at three different values of the shape parameter $a$.
      (a) Noise-induced synchronization ($a=1$). (b) Synchronization
      induced by intermediate impulses ($a=30$). (c) Phase locking
      ($a=1000$).}
    \label{fig.9}
  \end{center}
\end{figure}

Fluctuations of the Lyapunov exponent manifests itself in the dynamics of the phase difference.  In Fig.~\ref{fig.9}, typical time sequences of the phase difference $\Delta_{n}$ between the two oscillators in the synchronized states are shown. Three different shape parameters ($a=1,30,1000$) yielding nearly the same Lyapunov exponent $\langle \Lambda \rangle$ are chosen (with $\omega = 0.91$ fixed).
In noise-induced synchronization (Fig.~\ref{fig.9}(a)) with small $a$, the phase difference often exhibits big bursts, because the growth rate fluctuates strongly (large $\textrm{var}(\Lambda)$).  As the parameter $a$ increases, variation of the phase difference becomes smaller (Fig.~\ref{fig.9}(b)) and, in the phase-locked state, the phase difference stays small and rarely exhibits large bursts (Fig.~\ref{fig.9}(c)), because fluctuations of $\Lambda$ are very weak (small $\textrm{var}(\Lambda)$).  The peculiar dynamics shown in Fig.~\ref{fig.9}(a) is a direct consequence of the fluctuations of the linear growth rates and intimately related to the well-known on-off (or modulational) intermittency~\cite{Pikovsky,Pikovsky2,Fujisaka}.

\begin{figure}[tbhp]
  \begin{center}
    \includegraphics[width=0.75\hsize]{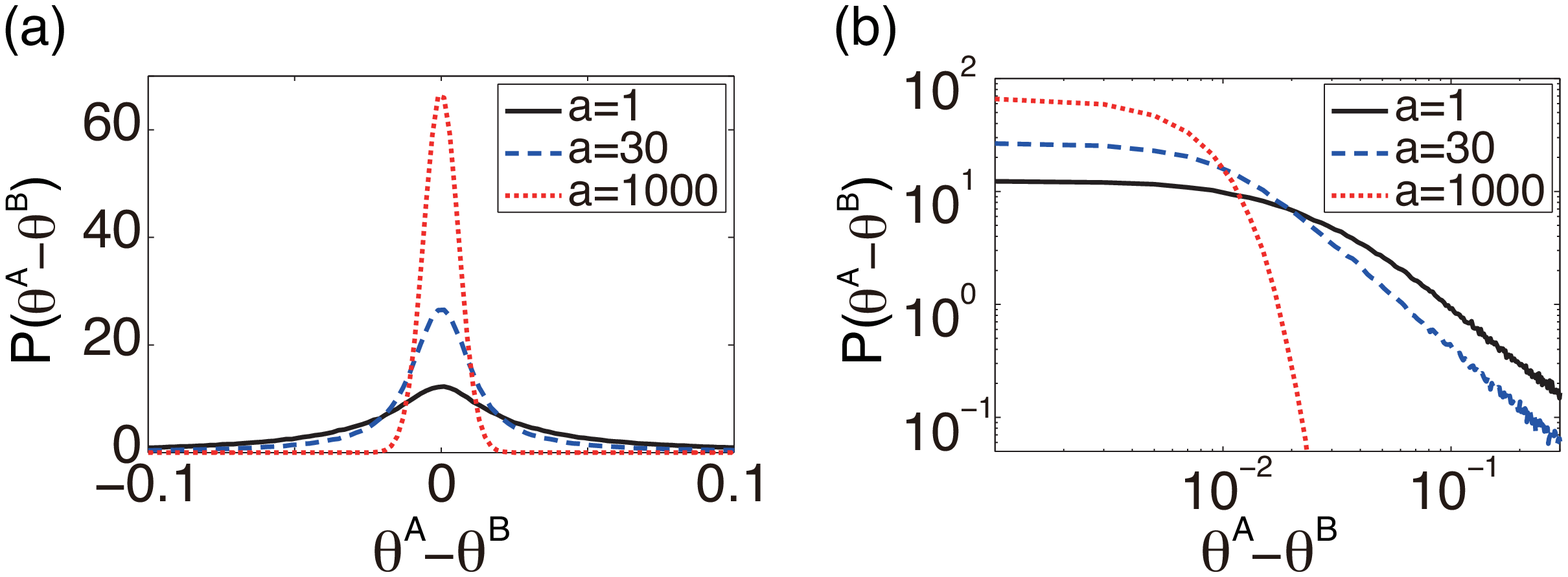}
    \caption{(Color online) Stationary probability distribution $P(\theta^A-\theta^B)$ of the phase difference $\theta^A-\theta^B$ of two oscillator
    calculated by direct numerical simulations of the random phase map.
    Black solid line, blue dashed line and red dotted line plot the results for $a=1,30$ and $1000$, respectively.
    (a)$P(\theta^A-\theta^B)$. (b)$P(\theta^A-\theta^B)$ using logarithmic scaling on the vertical axis.
    }
    \label{fig.10}
  \end{center}
\end{figure}

Figure~\ref{fig.10} shows the stationary PDFs of phase differences between the oscillators for differing values of the shape parameter $a$ in normal scales (a) and in log-log scales (b).  Reflecting the degree of the fluctuations of $\Lambda$, the width of the PDFs varies widely.  The tail part of the distribution exhibits power-law decay, which is broader for smaller $a$, indicating that the phase differences exhibit big bursts (phase differences) much more frequently.
As shown in Ref.~\cite{Pikovsky,Pikovsky2,Fujisaka,Kuramoto-Nakao}, the exponent of the power-law tail is approximately given by the ratio $\langle \Lambda \rangle / \mbox{var}( \Lambda ) $ that we plotted Fig.~\ref{fig.7}(c).  Thus, degree of intermittency in the synchronization dynamics also characterizes the difference in noise-induced synchronization and phase locking clearly.


\section{Information analysis}

We have observed that the joint PDFs of the oscillator phases and the
impulse intervals exhibit distinct characteristics depending on the
shape parameter.  In this section, to quantify correlations between
the oscillator phases and the impulse intervals for different types of
synchronization dynamics, we introduce information measures.  In
particular, we focus on the mutual dependence or
``causality'' of a pair of phase oscillators driven by a common
impulse source.  This leads us to physically distinct interpretations of the two synchronization scenarios.

\subsection{Information measures}

\subsubsection{Mutual information}

The mutual information $J(X;Y)$~\cite{Cover} is a nonnegative measure
that quantifies mutual dependence of two random variables $X$ and
$Y$~\cite{Cover}.  It is defined as
\begin{equation}
  J(X;Y)=\sum_{x\in X, y\in Y} p(x,y)\ln \frac{p(x,y)}{p(x)p(y)},
\end{equation}
where $p(x,y)$ is a joint probability distribution of $X$ and $Y$, and
$p(x)$ and $p(y)$ are probability distributions of $X$ and $Y$,
respectively. We often normalize $J(X,Y)$ by its upper limit as
\begin{equation}
  j(X;Y) = \frac{J(X;Y)}{\textrm{min}\{H(X), H(Y)\}},
\end{equation}
where $H(X)$ and $H(Y)$ are entropies of $X$ and $Y$, respectively. The normalized mutual information satisfies $0 \leq j(X;Y) \leq 1$.
Assuming, for example, that $X$ represents the inter-impulse
interval $\tau$ and $Y$ the oscillator phase $\theta$ just after this
interval, $J(X;Y)$ or $j(X;Y)$ quantifies how much the oscillator phase tells us about the impulses.

\subsubsection{Interaction information}

The interaction information $I(X;Y;Z)$~\cite{McGill,Tsujishita} is a
generalization of the mutual information $J(X;Y)$ to three random
variables $X$, $Y$, and $Z$, and quantifies mutual dependence among
them. It is defined as
\begin{equation}
  I(X;Y;Z) = \sum_{x\in X, y\in Y} p(x,y) \ln \frac{p(x,y)}{p(x)p(y)} - \sum_{x\in X, y\in Y, z\in Z} p(x,y,z) \ln \frac{p(x,y,z)p(z)}{p(x,z)p(y,z)},
\end{equation}
where $p(x,y,z)$ is a joint probability distribution function of $X$,
$Y$, and $Z$.  $I(X;Y;Z)$ is symmetric with respect to permutations of
$X$, $Y$, and $Z$.  The first term is simply the mutual information $J(X;Y)$
between $X$ and $Y$, and the second term on the right hand side gives
conditional mutual information $J(X;Y|Z)$ between $X$ and $Y$ given
$Z$.  Thus, $I(X;Y;Z)$ can be expressed as
\begin{equation}
  I(X;Y;Z) = J(X;Y) - J(X;Y|Z),
\end{equation}
which provides us with some intuitive meaning of $I(X;Y;Z)$. Namely, it
measures the effect of knowing $Z$ in guessing the mutual dependence
of $X$ and $Y$.

\begin{figure}[tbhp]
  \begin{center}
    \includegraphics[width=0.5\hsize]{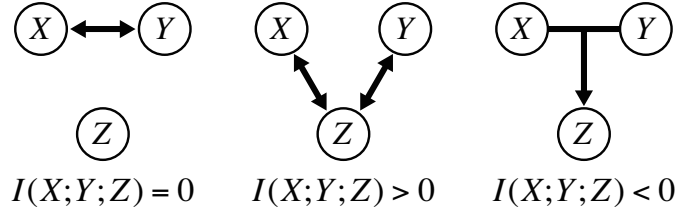}
    \caption{(Color online) Dependence diagram implied by the interaction information $I(X;Y;Z)$. Each arrow indicates the dependence between the variables.}
    \label{fig.11}
  \end{center}
\end{figure}

Unlike $J(X;Y)$, $I(X;Y;Z)$ can take both positive and negative
values.  The sign of $I(X;Y;Z)$ tells us about how the three random
variables $X$, $Y$, and $Z$ depend on each other, as illustrated in
Fig.~\ref{fig.11}.

(i) If $I(X;Y;Z)=0$, at least one of the three random variables is
independent of the others. For example, knowing $Z$ has no effect in
guessing the mutual information between $X$ and $Y$, $J(X;Y|Z) =
J(X;Y)$.  All variables are mutually dependent if $I(X;Y;Z) \neq 0$.

(ii) If $I(X;Y;Z)>0$, one of the three variables tends to determine the
other two variables. For example, knowing $Z$ lowers the information
on $X$ gained by observing $Y$ (or vice versa), i.e., $J(X;Y|Z) <
J(X;Y)$.  In particular, the maximum value of $I(X;Y;Z)$, given by
$\textrm{min}\{H(X), H(Y), H(Z)\}$, is attained when one of the three
variables dominates the other two, for example, as $X = h_X(Z)$ and $Y
= h_Y(Z)$ where $h_X$ and $h_Y$ are some maps.

(iii) if $I(X;Y;Z)<0$, knowing e.g. $Z$ helps in gaining information on
$X$ by observing $Y$, so that $J(X;Y|Z) > J(X;Y)$.  In particular,
$I(X;Y;Z)$ takes its maximum value when two of the random variables
dominate the remaining one, for example, when $Z = h(X, Y)$ with some
map $h$ and moreover $J(Y;Z) = 0$ ($Y$ and $Z$ are independent).  The
maximum value is equal to the negative conditional entropy $-H(Z|X)$
of the dominated variable $Z$ given the dominating variable $X$.

We normalize the interaction information $I(X;Y;Z)$ by its maximum as
\begin{equation}
  i(X;Y;Z) = \frac{I(X;Y;Z)}{\textrm{min}\{H(X),H(Y),H(Z)\}},
\end{equation}
so that $0 \leq i(X;Y;Z) \leq 1$ is satisfied.  In addition to the
above normalization, $I(X;Y;Z)$ may also be
normalized by the mutual information $J(X;Y)$ as
\begin{equation}
  i'(X;Y;Z) = \frac{I(X;Y;Z)}{J(X;Y)} = 1-\frac{J(X;Y|Z)}{J(X;Y)}.
\end{equation}
The quantity $i'(X;Y;Z)$ satisfies $0 \leq i'(X;Y;Z) \leq 1$ (because $I(X;Y;Z) \geq 0$ in the present case as shown below), quantifies how much information $Z$ conveys about the relationship between $X$ and $Y$.
We use these normalized $i(X;Y;Z)$ and $i'(X;Y;Z)$ to quantify mutual dependence among $\tau$, $\theta_{1}$, and $\theta_{2}$.


\subsection{Numerical results}

The mutual information and the interaction information are calculated from the PDFs obtained by numerically solving the Frobenius-Perron equations (\ref{eq:FP1}) and (\ref{eq:FP2}).  We discretize the inter-impulse intervals and the two oscillator phases using $100$ grid points and use the resulting coarse-grained discrete probability distributions in the calculation.
Note that we need the joint probability distribution $p(\theta^A, \theta^B, \tau)$ to calculate the interaction information, which corresponds to $100^3$
grid points.  Such a calculation is only feasible by the Frobenius-Perron approach.

\subsubsection{Mutual information $J(\tau ; \theta)$}

\begin{figure}[tbhp]
  \begin{center}
    \includegraphics[width=0.75\hsize]{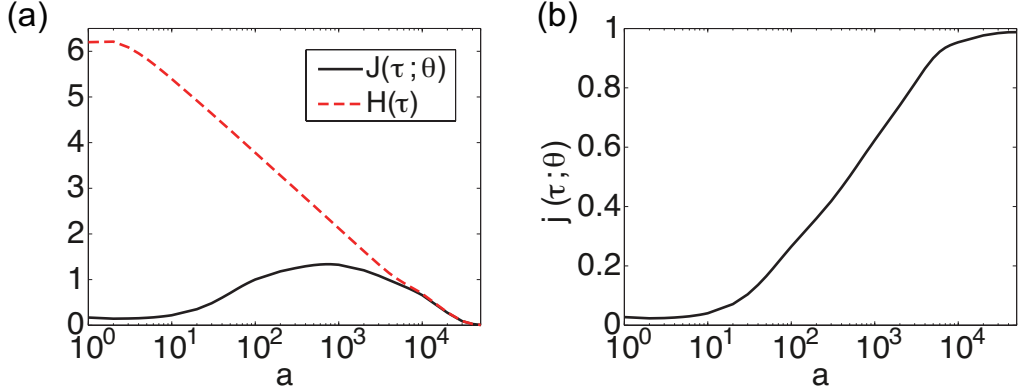}
    \caption{(Color online) (a) The mutual information $J(\theta;
      \tau)$ (black solid line) and $H(\tau)$ (red dashed line).  (b)
      Normalized mutual information $j(\theta; \tau)=J(\theta;
      \tau)/H(\tau)$.}
    \label{fig.12}
  \end{center}
\end{figure}

Figure~\ref{fig.12}(a) shows the mutual information $J(\tau ; \theta)$
of the inter-impulse interval $\tau$ and the phase $\theta$ just after
this interval, as well as its upper limit $H(\tau)$ (entropy of the
inter-impulse intervals) as functions of the shape parameter $a$ for
the resonant situation, $T = 1/\omega = \langle \tau \rangle = 1$.
$H(\tau)$ takes its maximum value in the Poisson limit ($a=1$),
monotonously decreases as the shape parameter $a$ is increased, and
almost vanishes for nearly periodic impulses ($a=10^5$).  The raw
mutual information $J(\tau ; \theta)$ takes a small value for Poisson
impulses ($a=1$) and gradually increases with the shape parameter
$a$. $J(\tau ; \theta)$ takes its maximum value at $a = a_\textrm{M}
\simeq 1000$, then decreases again, and almost vanishes when the
impulses become nearly periodic ($a=10^5$).
 
Figure~\ref{fig.12}(b) shows the normalized mutual information $j(\tau
; \theta) = J(\tau ; \theta) / H(\tau)$.  It increases smoothly from a
small value (nearly $0$) to $1$ as the shape parameter $a$ is
increased from $a=1$ (Poisson) to $a=10^5$ (nearly periodic).  This
indicates that the oscillator phase has only little dependence on the
inter-impulse interval just before it is measured in the Poisson case, whereas the oscillator phase possesses almost complete information about the interval in the periodic case.  These results are consistent with the shape of the single-oscillator phase PDF shown in Fig.~\ref{fig.3}.

It is interesting to note that the raw mutual information $J(\tau ; \theta)$ has a peak at the intermediate shape parameter $a$.
If we regard our model as an information channel, the output oscillator phase conveys information about the input impulse interval most efficiently at this value.  This can be interpreted as follows.  For periodic impulses,  even though $\theta$ is locked to $\tau$, the interval $\tau$ has no information because $H(\tau) = 0$ and therefore $J(\tau ; \theta) = 0$.  For Poisson impulses, though $H(\tau)$ is relatively large, $\theta$ does not faithfully represent $\tau$.  As a compromise, $J(\tau ; \theta)$ is maximized at the intermediate value of $a$.


\subsubsection{Interaction information $I(\tau ; \theta^A ; \theta^B)$}

\begin{figure}[tbhp]
  \begin{center}
    \includegraphics[width=0.75\hsize]{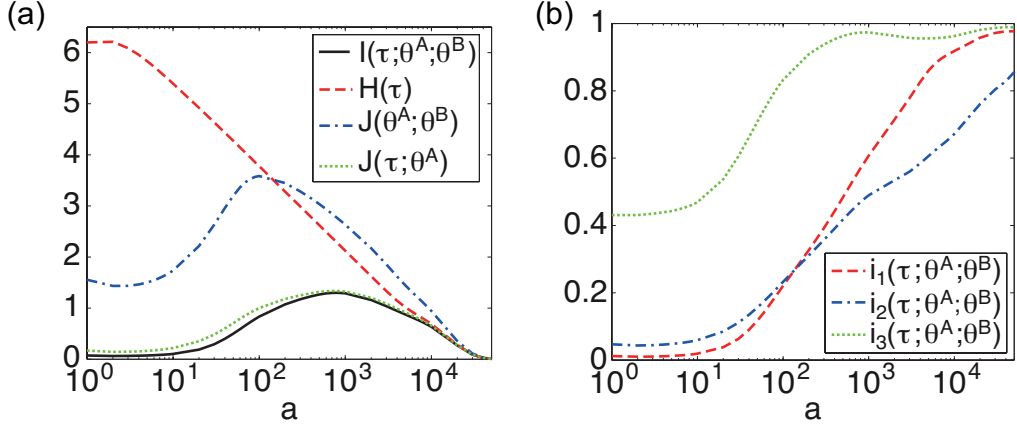}
    \caption{(Color online) (a) Interaction information $I(\theta^A;
      \theta^B; \tau)$ (black solid line) and normalization factors
      $H(\tau)$ (red dashed curve), $J(\theta^A ; \theta^B)$ (blue
      dot-dashed curve) and $J(\tau ; \theta^A)$ (green dotted curve)
      plotted against the shape parameter $a$.  (b) Normalized
      interaction information. $i_1(\theta^A; \theta^B;
      \tau)=I(\theta^A; \theta^B; \tau)/H(\tau)$ (red dashed curve),
      $i_2(\theta^A; \theta^B; \tau)=I(\theta^A; \theta^B;
      \tau)/J(\theta^A ; \theta^B)$ (blue dot-dashed curve), and
      $i_3(\theta^A; \theta^B; \tau)=I(\theta^A; \theta^B;
      \tau)/J(\tau ; \theta^A)$ (green dotted curve) }
    \label{fig.13}
  \end{center}
\end{figure}

More interesting insight can be attained by examining the interaction
information $I(\tau ; \theta^A ; \theta^B)$ among the inter-impulse
interval and the phases of the two driven oscillators.
Figure~\ref{fig.13}(a) shows the interaction information $I(\tau ;
\theta^A ; \theta^B)$, its upper limit $H(\tau)$, and mutual
informations $J(\theta^A ; \theta^B)$ and $J(\tau ; \theta^A)$ as
functions of the shape parameter $a$ for $\omega = 1$ (resonant).
The intermediate peak of the raw interaction information arises due to the same reason as in the previous case of the raw mutual information.
Interaction informations normalized in three ways, $i_1(\tau ;
\theta^A ; \theta^B) = I(\tau ; \theta^A ; \theta^B) / H(\tau)$,
$i_2(\tau ; \theta^A ; \theta^B) = I(\tau ; \theta^A ; \theta^B) /
J(\theta^A; \theta^B)$, and $i_3(\tau ; \theta^A ; \theta^B) = I(\tau
; \theta^A ; \theta^B) / J(\tau ; \theta^A)$, are plotted against $a$
in Fig.~\ref{fig.13}(b).

The normalized interaction information $i_1(\tau ; \theta^A ;
\theta^B)$ decreases as the shape parameter is decreased from
$a=50000$ (nearly periodic) to $a=1$ (Poisson).  For nearly periodic
impulses, $i_1(\tau ; \theta^A ; \theta^B)$ is nearly $1$, implying
that $\tau$, $\theta^A$, and $\theta^B$ are significantly dependent on
each other in the phase-locking regime.  As $a$ is decreased,
$i_1(\tau ; \theta^A ; \theta^B)$ gradually decreases and, for Poisson
impulses, $i_1(\tau ; \theta^A ; \theta^B)$ almost vanishes,
indicating that at least one of the $\theta^A$, $\theta^B$, and $\tau$
is almost independent of the others.

The normalized interaction information $i_2(\tau ; \theta^A ;
\theta^B)$ also decreases as the shape parameter is decreased from
$a=50000$ (nearly periodic) to $a=1$ (Poisson).  For nearly periodic
impulses ($a=50000$), $i_2(\tau ; \theta^A ; \theta^B)$ is nearly $1$,
which implies that the inter-impulse interval $\tau$ dominates the
oscillator phases $\theta^A$ and $\theta^B$.  As $a$ is decreased,
$i_2(\tau ; \theta^A ; \theta^B)$ decreases gradually, and for Poisson
impulses, it becomes very close to $0$, namely, $\tau$ is almost
independent of $\theta^A$ and $\theta^B$ and thus $\tau$ tells very
little about $\theta^A$ and $\theta^B$.

The last normalized interaction information $i_3(\tau ; \theta^A ;
\theta^B)$ also decreases as the shape parameter is decreased from
$a=50000$ (nearly periodic) to $a=1$ (Poisson).  For nearly periodic
impulses ($a=50000$), $i_3(\tau ; \theta^A ; \theta^B)$ is nearly $1$,
which indicates that the phase $\theta^B$ has almost complete
information about $\tau$ and $\theta^A$.  It means that all $\tau$,
$\theta^A$ and $\theta^B$ depend on each other strongly.  Unlike
$i_2(\tau ; \theta^A ; \theta^B)$, $i_3(\tau ; \theta^A ; \theta^B)$
takes relatively large values around $0.4$ even in the Poisson limit,
which implies that we can get nearly $50\%$ of the information about
$\tau$ and $\theta^A$ just by looking at $\theta^B$.  This is because
the oscillators are synchronized even if they are driven by Poisson
impulses, and we can learn much about $\theta^A$ by knowing
$\theta^B$.

Though the interaction information itself is symmetric with respect to
its three arguments, we can obtain further insight by taking into
account the drive-response configuration of our model, namely, the
fact that the inter-impulse interval $\tau$ has a distinct meaning
from the other two phase variables $\theta^A$ and $\theta^B$.
Therefore, for the phase-locked situation with $i_1(\tau ; \theta^A ;
\theta^B) \simeq 1$, we may consider that the phases $\theta^A$ and
$\theta^B$ are predominantly controlled by the driving impulse $\tau$,
as implied from the condition giving the maximum value of $I(\tau ;
\theta^A ; \theta^B)$.  On the other hand, in the Poisson case with
$i_1(\tau ; \theta^A ; \theta^B)$ close to $0$, we may conclude that
$\tau$ is nearly independent of $\theta^A$ and $\theta^B$.  This
indicates that the phases have very little information about the
driving impulse in the noise-induced synchronized state.

\begin{figure}[tbhp]
  \begin{center}
    \includegraphics[width=0.5\hsize]{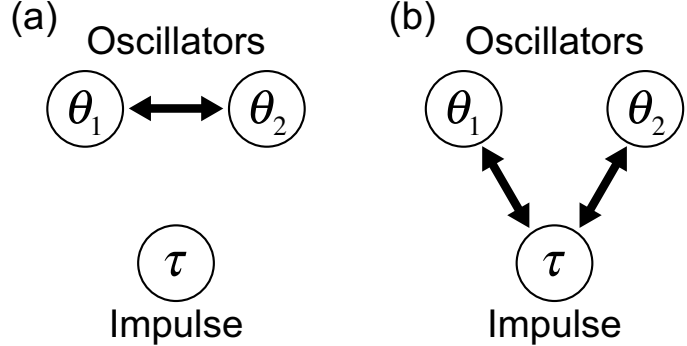}
    \caption{Schematic diagrams showing mutual dependence among the
      impulse interval and the oscillator phases inferred from the
      interaction information.  (a) Noise-induced synchronization.
      (b) Phase locking. }
    \label{fig.14}
  \end{center}
\end{figure}

The above results lead us to two distinct interpretations of the two synchronization scenarios, as schematically summarized in Fig.~\ref{fig.14}; in the phase locking, the two oscillators are simply oscillators dominated by the impulses, whereas they are almost free from the impulses but still behave synchronously in the noise-induced synchronization.


\section{Summary}

In the present study, we analyzed uncoupled oscillators driven by gamma impulses that smoothly interpolate between periodic and Poisson inter-impulse intervals to quantify the difference in synchronization dynamics between these two limiting cases.  By examining the dependence of various quantities on the shape parameter of the gamma impulses, we have revealed a clear difference between the two types of synchronization.  Using phase distributions, frequency detuning, statistics of Lyapunov exponents, and information-theoretic measures, we have quantitatively confirmed our original intuition, namely, that the oscillators are principally dominated by the impulses and mutually synchronized as its consequence for the periodic driving, whereas the oscillators are not entrained by the impulses (at least not strongly influenced by the inter-impulse interval just before the phase is measured) but they still attain coherence in the case of Poisson driving.


Generalization of our results for uncoupled oscillators to mutually interacting oscillators, in particular with delay, under common or correlated gamma impulses is an interesting future subject, where not only synchronization but qualitatively different behaviors~\cite{Lin} can be expected.
More detailed quantification of the input-output and the output-output relations of the impulse-driven oscillators from the viewpoint of information transfer would also be interesting, where effect of the functional shape of the phase response curve and its optimization would be important issues.
As we found in the measurement of mutual and interaction information, the transition between the phase locking and the noise-induced synchronization may not be simply monotonic from such a viewpoint, implying the importance of intermediate randomness of the input impulses.
We plan to push forward in these directions in our future studies.

\vspace{12pt}

We thank H. Suetani and H. Hata (Kagoshima), and Y. Tsubo and
J. Teramae (RIKEN) for illuminating comments.  This work is supported
by MEXT (Grant no.~22684020) and by the GCOE program ``The Next Generation
of Physics, Spun from Universality and Emergence'' from MEXT, Japan.


\section*{Appendix. Fourier representation of the Frobenius-Perron equation}

Here we introduce Fourier representation of the Frobenius-Perron
equations (\ref{eq:FP1}) and (\ref{eq:FP2}).  Based on this
representation, we can use the Fast Fourier Transform (FFT) algorithm
to numerically calculate convolution integrals included in the
equations and drastically reduce calculation costs.

Firstly, to numerically convolute $P_n$ and $S$, we transform the
Dirac delta function in the Frobenius-Perron equation, in such a way
that it becomes an explicit function of the variable of integration
$\theta'$ rather than the original phase variable $\theta$.
\begin{align}
  P_{n+1}(\theta)= & \sum_{k=-\infty}^{\infty} \int_0^1d\theta'
  \int_0^\infty d\tau \int_{-\infty}^{\infty}d\eta
  W(\tau)P_{n}(\theta')R(\eta)\delta \left( \theta - F(\theta'+\eta) - \omega \tau - k \right)\\
  = & \sum_{k=-\infty}^{\infty} \int_0^1d\theta' \int_0^\infty d\tau
  \int_{-\infty}^{\infty}d\eta
  W(\tau)P_{n}(\theta')R(\eta)\\
  & \qquad \qquad \qquad \qquad \times\frac{dF^{-1}(\theta-\omega\tau-k)}{d\theta}\delta
  \left( \theta'-F^{-1}(\theta-\omega\tau-k) + \eta \right),
  \label{FP01}
\end{align}
where we have defined an inverse function of the phase map,
\begin{equation}
  \mathcal{F}(x) = F^{-1}(x)\ \textrm{mod}\ 1.
\end{equation}
Since $F$ is a phase map,
\begin{align}
  F^{-1}(\theta-\omega\tau-k) = & F^{-1}(\theta-\omega\tau)-k =
  \mathcal{F}(\theta-\omega\tau)+d-k
\end{align}
holds for $F^{-1}(\theta-\omega\tau) \in [d, d+1)$, where $d$ is an
integer. Using this, we obtain
\begin{align}
  P_{n+1}(\theta) = & \sum_{k=-\infty}^{\infty} \int_0^1d\theta' \int_0^\infty d\tau W(\tau)P_{n}(\theta') \frac{d[\mathcal{F}(\theta-\omega\tau)+d-k]}{d\theta} R\left( \mathcal{F}(\theta-\omega\tau)+d-k-\theta' \right)\\
  = & \int_0^1 d\theta' \int_0^\infty d\tau W(\tau)P_{n}(\theta') \frac{d\mathcal{F}(\theta-\omega\tau)}{d\theta} \sum_{k=-\infty}^{\infty} R\left( \mathcal{F}(\theta-\omega\tau)-k-\theta' \right)\\
  = & \int_0^1d\theta' \int_0^\infty d\tau W(\tau)P_{n}(\theta')
  \frac{d\mathcal{F}(\theta-\omega\tau)}{d\theta} S\left(
    \mathcal{F}(\theta-\omega\tau)-\theta' \right),
\end{align}
where
\begin{align}
  S\left( \mathcal{F}(\theta-\omega\tau)-\theta' \right)=
  \sum_{k=-\infty}^{\infty} R\left(
    \mathcal{F}(\theta-\omega\tau)-k-\theta' \right).
\end{align}
Now, using the discrete Fourier transformation, we can convolute $P_n$
and $S$ as
\begin{align}
  P_{n+1}(\theta)= & \int_0^\infty d\tau W(\tau) \frac{d\mathcal{F}(\theta-\omega\tau)}{d\theta} \int_0^1d\theta' \sum_m p_{n,m} e^{2\pi im\theta'} \sum_l s_l e^{2\pi i l\left( \mathcal{F}(\theta-\omega\tau)-\theta'\right)}\\
  = & \int_0^\infty d\tau W(\tau) \frac{d\mathcal{F}(\theta-\omega\tau)}{d\theta} \sum_m p_{n,m} s_m e^{2\pi im\mathcal{F}(\theta-\omega\tau)}\\
  = & \sum_m p_{n,m} s_m \int_0^\infty d\tau W(\tau)
  D_m(\theta-\omega\tau),
\end{align}
where $p_{n,m}$ and $s_m$ are $m$th Fourier coefficients of $P_n$ and
$S$, respectively:
\begin{align}
  p_{n,m} = & \int_0^1 d\theta' P_n(\theta')e^{-2\pi m\theta'}\\
  s_m = & \int_0^1 d\theta' S(\theta')e^{-2\pi m\theta'},
\end{align}
and
\begin{align}
  D_m(\theta-\omega\tau) =
  \frac{d\mathcal{F}(\theta-\omega\tau)}{d\theta} e^{2\pi
    im\mathcal{F}(\theta-\omega\tau)}.
\end{align}

Next, we need to convolute the functions $W(\tau ; a,b) $ and
$D_m(\theta - \omega \tau)$.  From the definition of the gamma
distribution, Eq.~(\ref{eq:gamma}),
\begin{equation}
  W(\tau;a,b)d\tau = W(\omega\tau; a, \omega b)d(\omega\tau).
\end{equation}
holds.  Using this, we obtain
\begin{align}
  P_{n+1}(\theta)= & \sum_m p_{n,m} s_m \int_0^{\infty} d(\omega \tau) W(\omega\tau; a, \omega b) D_m(\theta-\omega\tau)\\
  = & \sum_m p_{n,m} s_m \int_0^{\infty} d(\omega \tau) \sum_l W_l e^{2\pi il\omega\tau} \sum_\alpha d_{m,\alpha} e^{2\pi i \alpha(\theta-\omega\tau)}\\
  = & \sum_{m, \alpha} p_{n,m}s_m d_{m,\alpha} W_\alpha e^{2\pi
    i\alpha \theta},
\end{align}
where $w_\alpha$ and $d_{m,\alpha}$ are $\alpha$th Fourier
coefficients of $W(\omega\tau; a, \omega b)$ and $D_m$, respectively:
\begin{align}
  W_\alpha = & \int_0^\infty dt W(t; a, \omega b) e^{-2\pi i \alpha t}\\
  d_{m,\alpha} = & \int_0^1 d\theta D_m(\theta)e^{-2\pi i \alpha
    \theta}.
\end{align}
Therefore, we can write the evolution of Fourier coefficient of
$P_n(\theta)$ as
\begin{equation}
  p_{n+1, \alpha} = \sum_m p_{n, m} s_m d_{m,\alpha} W_{\alpha},
\end{equation}
which can easily be calculated numerically.

Similarly, we can write the evolution of $p_{n, m, l}$, which is the
$(m, l)$th Fourier coefficient of the joint probability density
function $P_n(\theta^A, \theta^B)$,as
\begin{equation}
  p_{n+1, \alpha, \beta} = \sum_{m,l} p_{n, m, l} s_m s_l d_{m,\alpha} d_{l, \beta} W_{\alpha+\beta},
\end{equation}
which is also easy to calculate numerically.


\begin{thebibliography}{99}

\bibitem{Uchida-Roy} R. Roy and K.S. Thornburg, Jr.,
  Phys. Rev. Lett. {\bf 72}, 2009 (1994); A. Uchida, R. McAllister,
  and R. Roy, Phys.  Rev.  Lett. {\bf 93}, 244102 (2004); A. Uchida,
  K. Yoshimura, P. Davis, S. Yoshimori, and R. Roy, Phys. Rev. E {\bf
    78}, 036203 (2008).

\bibitem{Yoshida} K. Yoshida, K. Sato, A.  Sugamaga, J. Sound and
  Vibration {\bf 290}, 34 (2006).

\bibitem{Arai-Nakao} K. Arai and H. Nakao, Phys. Rev. E {\bf 77},
  036218 (2008).
  
\bibitem{Nagai-Nakao} K. Nagai and H. Nakao Phys. Rev. E {\bf 79},
  036205 (2009).
  
\bibitem{Mainen-Sejnowski} Z. F. Mainen and T. J. Sejnowski, Science
  {\bf 268}, 1503 (1995).

\bibitem{Binder-Powers} M. D. Binder and R. K. Powers, Journal of
  Neurophysiology {\bf 86}, 2266 (2001).

\bibitem{Galan} R. F. Gal\'an, N. F. Trocme, G. B. Ermentrout, and
  N. N. Urban, J. Neurosci {\bf 26(14)}, 3646 (2006).

\bibitem{Danzl} P. Danzl, R. Hansen, G. Bonnet, and J. Moehlis, Journal of Computational Neuroscience, {\bf 25}, 141 (2008).

\bibitem{Ermentrout-Galan-Urban} G. B. Ermentrout, R. F. Gal\'an, and
  N. N. Urban, Trends Neurosci. {\bf 31}, 428 (2008).

\bibitem{Royama} T. Royama, {\it Analytical population dynamics}
  (Chapman and Hall, London, UK, 1992).

\bibitem{Ranta} E. Ranta, V. Kaitala and E. Helle, Oikos \textbf{78}, 136 (1997).

\bibitem{Koenig} W. D. Koenig and J. M. H. Knops, Nature \textbf{396}, 225 (1998).

\bibitem{Moran} P. A. P. Moran, Aust. J. Zool. \textbf{1}, 291 (1953).

\bibitem{Winfree} A. T. Winfree, {\it The Geometry of Biological Time}
  (Springer-Verlag, New York, 2001).

\bibitem{Kuramoto} Y. Kuramoto, {\it Chemical Oscillation, Waves, and
    Turbulence} (Springer-Verlag, Tokyo, 1984) (republished by Dover,
  New York, 2003).

\bibitem{Pikovsky} A. Pikovsky, M. Rosenblum, and J. Kurths, {\it
    Synchronization} (Cambridge University Press, England, 2001).

\bibitem{Glass} L. Glass and M. C. Mackey, {\it From Clocks to Chaos:
    The Rhythms of Life} (Princeton University Press, USA, 1988).

\bibitem{Yamanobe} T. Yamanobe and K. Pakdaman,
  Biol. Cybern. \textbf{86}, 155 (2002).

\bibitem{Nakao-Arai} H. Nakao, K. Arai, K. Nagai, Y. Tsubo, and
  Y. Kuramoto, Phys. Rev. E {\bf 72}, 026220 (2005).
  
\bibitem{Galan-Ermentrout-Urban} R. F. Gal\'an, G. B. Ermentrout and
  N. N. Urban, Phys. Rev. Lett. {\bf 94}, 158101 (2005).
  
\bibitem{Tateno-Robinson} T. Tateno and H. P. C. Robinson, Biophysical
  Journal {\bf 92}, 683 (2007).

\bibitem{Arai-Nakao2} K. Arai and H. Nakao, Phys. Rev. E {\bf 78},
  066220 (2008).

\bibitem{Hansel}
  D. Hansel, G. Mato, and C. Meunier, Neural Computation {\bf 7}, 307 (1995).

\bibitem{Abouzeid-Ermentrout} A. Abouzeid and G. B. Ermentrout,
  Phys. Rev. E {\bf 80}, 011911 (2009).

\bibitem{Marella-Ermentrout} S. Marella and G. B. Ermentrout,
  Phys. Rev. E {\bf 77}, 041918 (2008).
  
\bibitem{Brown} E. Brown, J. Moehlis, and P. Holmes, Neural
  Computation {\bf 16}, 673 (2004).

\bibitem{Izhikevich} E. Izhikevich, {\it Dynamical Systems in
    Neuroscience. The Geometry of Excitability and Bursting.} The MIT
  Press, 2007.
  
\bibitem{Ermentrout-typeI} G. B. Ermentrout, Neural Computation {\bf 8}, 979 (1996).

\bibitem{Shimokawa} T. Shimokawa, S. Koyama and S. Shinomoto, J. Comp. Neurosci. DOI: 10.1007/s10827-009-0194-y (2010).

\bibitem{Santos} G. J. Escalera Santos and P. Parmananda, Phys. Rev. E, {\bf 65}, 067203 (2002).

\bibitem{Doi} S. Doi, J. Inoue and S. Kumagai,
  J. Stat. Phys. \textbf{90}, 1107 (1998).

\bibitem{Lasota-Mackey} A. Lasota and M. C. Mackey, {\it Probabilistic
    properties of deterministic systems} (Cambridge University Press,
  Cambridge, 1985).

\bibitem{Ott} E. Ott, {\it Chaos in Dynamical Systems} (Cambridge
  University Press, Cambridge, 2002).

\bibitem{Ermentrout-Saunders} G. B. Ermentrout and D. Saunders,
  J. Comput. Neurosci. {\bf 20}, 179 (2006).

\bibitem{Yoshimura-Arai-Davis} K. Yoshimura, P. Davis, and A. Uchida, in Proceedings of the 2007 International Symposium on Nonlinear Theory and its Applications, Vancouver, Canada (IEICE, Vancouver, 2007), p. 104-107.

\bibitem{Fujisaka} H. Fujisaka and T. Yamada, Prog. Theor.  Phys. {\bf
    69}, 32 (1983); H. Fujisaka, Prog. Theor. Phys. {\bf 70}, 1264
  (1983); H. Fujisaka and T. Yamada, Prog. of Theor.  Phys. {\bf 74},
  918 (1985).

\bibitem{Pikovsky2} A. S. Pikovsky, Phys. Lett. A {\bf 165}, 33 (1992).

\bibitem{Kuramoto-Nakao} Y. Kuramoto and H. Nakao, Phys. Rev. Lett. {\bf 78}, 4039 (1997).

\bibitem{Cover} T. A. Cover and J. A. Thomas, {\it Elements of Information Theory} (Wiley, 1991).

\bibitem{McGill} W. J. McGill, Psychometrika, \textbf{19}, 97 (1954).

\bibitem{Tsujishita} T. Tsujishita, Advances in Applied Mathematics {\bf 16}, 269 (1995).

\bibitem{Lin} K. K. Lin, E. Shea-Brown, and L.-S. Young, Journal of
  Nonlinear Science {\bf 19}, 1432 (2009).
  
  \end{thebibliography}
\end{document}